\documentclass[iop]{emulateapj} \pdfoutput=1
\usepackage{amsmath}
\newcommand\sB{{\textbf {\em B}}}

\newcommand\sV{{\textbf {\em V}}}

\newcommand\divb{{\nabla \! \cdot \! \textbf{\em B}}}

\newcommand\gadget{{\sc GADGET-2}}
\newcommand\jcoph{{J.~Comp.~Phys.}}

\begin{document}

\title{Phurbas: An Adaptive, Lagrangian, Meshless,
  Magnetohydrodynamics Code. II. Implementation and Tests}
\shorttitle{Phurbas MHD Code. II. Tests}

\author{Colin P.~M\textsuperscript{c}Nally\altaffilmark{2}}
\email{{\tt cmcnally@amnh.org}}
\author{Jason L.~Maron\altaffilmark{1}}
\email{{\tt jmaron@amnh.org}}
\author{Mordecai-Mark Mac Low\altaffilmark{2}}
\affil{Department of Astrophysics, American Museum of Natural History New York, NY, USA}
\email{{\tt mordecai@amnh.org}}

\altaffiltext{1}{current address: North Carolina Museum of Natural
  Sciences, Raleigh, NC, USA}
\altaffiltext{2}{Department of Astronomy, Columbia University, New
  York, NY, USA}

\begin{abstract} 

We present an algorithm for simulating the equations of ideal
magnetohydrodynamics and other systems of differential equations on
an unstructured set of points represented by sample particles.
The particles
move with the fluid, so the time 
step is not limited by the Eulerian Courant-Friedrichs-Lewy
condition. Full spatial adaptivity is
required to ensure the particles fill the computational volume, 
and gives the algorithm substantial
flexibility and power.  A target resolution is specified for each point in
space, with particles being added and deleted as needed to meet this target.
We have parallelized the code by adapting the framework provided by
\gadget{}.  A set of standard test
problems, including $10^{-6}$ amplitude linear MHD waves, magnetized
shock tubes, and Kelvin-Helmholtz instabilities is presented. 
Finally we demonstrate good agreement with analytic
predictions of linear growth rates for magnetorotational instability
in a cylindrical geometry.
This paper documents the Phurbas algorithm as implemented in Phurbas version 1.1.
\end{abstract}


\section{Introduction}

In \citet[][hereafter Paper I]{phurbasalg} we described an adaptive,
Lagrangian, meshless, method for magnetohydrodynamics (MHD).  We now
describe the parallel implementation of this algorithm and its tests,
and discuss its practical properties.  The test problems used are
selected from the accepted ones in the literature; many follow
those documented for Athena\footnote{\tt https://trac.princeton.edu/Athena/}
\citep{2008ApJS..178..137S}, and those used by
\cite{2000JCoPh.161..605T}, and \cite{1995ApJ...442..228R}.  The three
goals of the tests are to verify the convergence to smooth
solutions of the MHD equations, demonstrate the global conservation
properties and shock errors, and finally to verify the ability of
Phurbas to model the linear growth of magnetorotational instability correctly.

In \S~\ref{sec_implementation} we describe how we have implemented our
algorithm into a parallel code called Phurbas using the \gadget{} 
code\footnote{\tt  http://www.mpa-garching.mpg.de/gadget/} 
\citep{2005MNRAS.364.1105S} as a basis.  We describe the tests
that we have performed with this implementation in \S~\ref{sec_tests}.
We summarize the results of the tests, and discuss the implications
and future prospects for Phurbas-like methods in
\S~\ref{sec_discussion}.

\section{Implementation of the Algorithm}

Phurbas is a parallel code using the Message Passing Interface,
following the patterns of the \gadget{} code
\citep{2005MNRAS.364.1105S}, which originally combined a smoothed
particle hydrodynamics (SPH) treatment of gas dynamics with a tree
solution of the Poisson equation to follow N-body dynamics. The serial
particle update module described in Paper~I has been incorporated into
a modified version of \gadget{} that has been re-purposed to serve as
a parallel framework.  We refer to the version documented in this work
as Phurbas version 1.1 to allow for the differentiation of future
modifications to the algorithm and the code.  (We note that tests of Phurbas
1.0 were described in the first submitted version of this paper, posted
to ArXiv as version 1 of this paper.) 

We summarize here the algorithm described in detail in Paper I.
Phurbas solves the equations of ideal MHD, expressed in terms of
Lagrangian time derivatives.  The equations are discretized on an
adaptive set of particles that carry values of the field variables
(density $\rho$, velocity $\sV$, magnetic field $\sB$, and internal
energy density $\sigma$).  The particles move with the velocity $\sV$
that they carry, making the Lagrangian formalism the natural
description of the time evolution of the field variables. The
evolution equations relate time derivatives of the field variables to
their spatial derivatives. To calculate the spatial derivatives,
Phurbas uses a local, third-order, polynomial, moving least squares
interpolation, using
neighbor particle values drawn from a sphere of radius
$r_{f,i}=2.3\lambda_i$ about particle $i$, where $\lambda_i$ is the
effective resolution parameter.  
A second order predictor-corrector scheme is used with the time
derivatives obtained by applying the MHD equations to the approximate 
spatial derivatives to advance the particle positions and variable values.

Particles are added and deleted, filling voids and destroying clumps, 
to ensure that each sphere of radius $r_f$ is well sampled.
On average, the particle creation and deletion results in 
at least one particle per volume $\lambda^3$.
The resolution parameter $\lambda$ need not be constant 
in space or time, and each particle has an individual time step.
In this sense Phurbas is fully adaptive both spatially and temporally.
The time steps are 
independent of the bulk velocity of the flow.
Spatial variation of the time steps is limited to prevent the 
penetration of short time step particles into regions 
of long time step particles.

To obtain numerical stability, Phurbas integrates a modified version of the ideal MHD
equations including a bulk viscosity. 
The bulk viscosity coefficient is broken into two scalar fields.
The first is scaled so that motions near the grid scale are always damped.
The second adapts to provide damping in regions with large change, 
such as shocks and contact discontinuities.
A diffusive correction term in the induction equation locally diffuses nonzero $\divb$ away.

\subsection{\gadget{} Based Implementation}\label{sec_implementation}

In this section we outline the changes made to \gadget{} to adapt it
to Phurbas, which serves also to highlight the differences in
the infrastructure needed to support SPH and Phurbas.  The largest
modifications are the removal of the SPH algorithm for evaluating the
spatial derivatives, the addition of data passing and elements in the
data structures needed for the Phurbas MHD algorithm, the modification
to support addition and deletion of gas-type particles, the addition
of magnetic field variables, and a new main time advance loop.
Additions to the input-output routines and global statistics were also made.

In \gadget{} SPH, the sums required for value and gradient evaluation
are computed in a distributed manner. The contribution to each
sum from a set of neighbor particles can be computed on the processor
where those neighbors reside, so that only the total value need be
communicated.  For the Phurbas particle update, the values for all
particles within $r_{f,i}$ of particle $i$ must be communicated to the
process hosting particle $i$.  Particles are updated in batches, so
within groups of $\sim 5,000$--10,000 particles duplicate neighbor
communications can be avoided.

Phurbas requires only a strict radius-based neighbor search, unlike in
\gadget{} SPH where a mutual neighbor relation is required to achieve
symmetric interparticle forces.  The Phurbas neighbor search is also
performed on a fixed radius volume, so it does not need iterative refinement
like the \gadget{} SPH neighbor search, which adapts the neighbor
search radius to include a target number of neighbors.  In Phurbas,
the moving least squares interpolation and 
the particle adaption algorithm require
information from the neighbor particles of particle $i$ to be
retrieved to the host process of particle $i$.  The set of particles
used for these three procedures is the same, the particles within
$r_f$.  

Compared to \gadget{} SPH particles, Phurbas particles use somewhat more
memory.  Phurbas uses 56.5 eight byte variables per
particle as opposed to the 36.5 used by \gadget{} in double precision mode.
We retain the basic structures of \gadget{} in that the
variables required for the neighbor tree construction are stored in a
separate array from the fluid quantities.  
As the memory required to complete the
retrieval of neighbor information can be much larger than in \gadget{}
SPH, and dynamically varies, we dynamically allocate buffers as needed
to complete this step and free the memory after the process is
finished.  Compared to \gadget{} SPH, we have modified the code to
avoid persistent allocation of memory, which in particular alleviates
problems with computing the domain decomposition with large numbers of
parallel processes.  

The void search in Phurbas described in Paper~I relies on computing
distances to neighbor particles placed on a $9^3$ grid.
This procedure gains in computational speed if the three-dimensional grid is
implemented as a one-dimensional list in memory that is shortened each time a grid
point is eliminated from consideration as a nearest neighbor. This minimizes the number of
times each grid point must be accessed, while keeping the values
arranged compactly in memory. We have implemented this strategy by simply replacing the
coordinates of any eliminated point with the coordinates of the
current last point on the list and shortening the length of the list.

\gadget{} does not support
dynamic particle addition and deletion.  In Phurbas, we first perform
a load balance on the list of particles to add to voids, in order to distribute them
among processes evenly. The particles are created in free memory
spaces on the processors where they are moved to by the load balance
algorithm. Particles in clumps are deleted from the processors they are resident
on.

As we have now deleted some particles, and added others to essentially
random processors, we perform a new global \gadget{} load balance calculation,
followed by a Peano-Hilbert ordering and tree build, as in the
standard \gadget{} algorithm \citep{2005MNRAS.364.1105S}.  Doing this
on the entire particle list brings the new particles to optimal
positions on the processors and provides neighbor information for the
subsequent processing stages.

The restriction of local time step variations only requires
information from particle $i$ to be sent to the host processes of the
neighbors of particle $i$.Time steps are rounded down to powers of two in order to use the
\gadget{} binary block time step scheme \citep{2005MNRAS.364.1105S},
and the increase in time step is limited with the \gadget{}
synchronized time step scheme.  

The most significant optimization made has been for the assembly of
the left hand side matrix for the least squares fitting problem used
in the polynomial interpolation.  This has been explicitly
loop-unrolled into very simple FORTRAN designed to maximize the
ability of the compiler to pipeline the instructions and optimize
cache use.  Despite the increased memory use compared to \gadget{}
SPH, our experience has been that simulations are computation limited
rather than memory limited.

Compared to dark matter dominated, cosmological, N-body problems, pure MHD
fluid problems intrinsically require more work per particle per time
step.  For the highest resolution run in the cylindrical geometry MRI
test in the following section, we used $\sim 10^6$
particles on 48
cores, 95\% of which were typically active, non-boundary
particles. This was run on the Texas Advanced Computation Center
Lonestar cluster, which consists of nodes with dual Xeon Intel,
hexacore, 3.33~GHz, 64-bit, Westmere processors (13.3 GFLOPS per core)
interconnected with 
InfiniBand QDR.  
Phurbas took an average
$\sim 1.3\times10^{-4}$ seconds for the serial procedures per particle
(void and clump checks,
moving least squares interpolations, time derivative
calculations, and time integration corrector step).  We note that the
performance of this section of the code depends highly on the compiler
and optimization settings used.  The wallclock time per step was $\sim
5$ seconds for a step involving 
$\sim 9.1 \times 10^5$ active particles, giving a speed of
$\sim 2.6\times10^{-4}$ seconds per particle or
$3.8\times10^3$ total updates per core per wallclock second, including
adaptivity involving $\sim 1500$ additions and deletions.  It should
be noted that Phurbas version 1.1 code has not yet been heavily optimized, so
we believe the parallel overhead can be further reduced.

For a given application to be suitable for simulation with Phurbas, the Lagrangian nature, 
adaptivity, and individual time steps of
Phurbas must offer a significant advantage on a problem. Otherwise
a high-order, grid-based method such as the Pencil Code
\citep{2002CoPhC.147..471B} is a more efficient choice, as it computes
more than $10^2$ times more updates per core per second.

\section{Tests} \label{sec_tests}

We present in this section a series of tests that serve to 
verify the performance of Phurbas, and illustrate its abilities and
limitations.  These include linear waves (Sec.~\ref{sec_linwaves}),
circularly polarized Alfv\'{e}n waves (Sec.~\ref{sec_circwaves}),
hydrodynamical and MHD shock
tubes (Sec.~\ref{sec_shocks}), Kelvin-Helmholtz instabilities
(Sec.~\ref{sec_khi}), and cylindrical magnetorotational instability
(Sec.~\ref{sec_flocktest}). 

As Phurbas is meshless, it is an intrinsically three-dimensional
algorithm. The performance and design criteria are different in different numbers of dimensions.
Hence, even when tests have symmetries such that they could be stated in a lower
number of dimensions, we perform them in fully three-dimensional domains
to achieve results reflective of the true capabilities of the code.

It is important to use realistic particle distributions for these tests.
Though perfectly gridded particle distributions would yield good
results for some tests, as has been demonstrated in the literature,
such well arranged distributions cannot be expected in a general flow.
Instead, we realize cubes of relaxed, glassy particle distributions using several iterations of
Phurbas's void and clump detection algorithms, and then tile these
distributions to fill the problem domain. 
This also ensures that initially, no particle adaption occurs until
the particles move far enough to justify it. 
This allows both the use of a realistic particle distribution and makes it possible to create a set of
predictably refined initial conditions for convergence studies.
Irregular initial particle distributions also remove the need to set up tests
at odd angles to prevent aligning waves and shock fronts with a grid.

\subsection{Linear Waves} \label{sec_linwaves}
\begin{figure}
\plotone{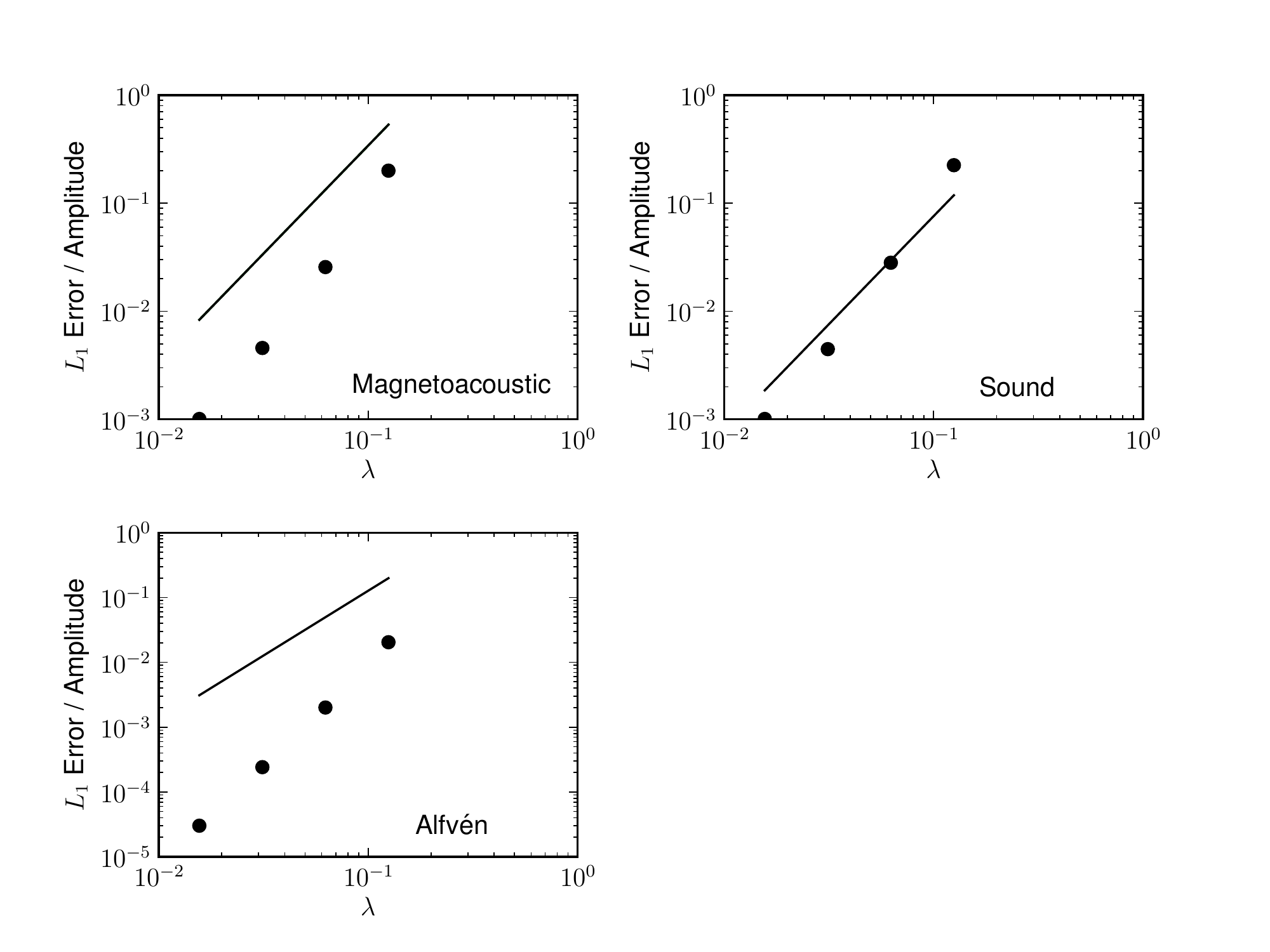}
\caption{Convergence of linear waves. Points show the relative $L_1$-norm errors
  derived from the comparison of Phurbas results to solutions of 
  the MHD dispersion relation including the stabilizing bulk viscosity
  (see Appendix~\ref{sec_dispersion}). The slopes plotted show
  second-order convergence.
}
\label{figlinwaveerr}
\end{figure}

To verify the convergence of the scheme and its accuracy in the linear
domain, three MHD waves are modeled.  The three waves are a
magnetoacoustic wave propagating perpendicularly to the background
magnetic field, 
a shear Alfv\'{e}n wave propagating along the background field,
and a sound wave propagating in an unmagnetized medium. 
We compare the results with
solutions of the dispersion relation
for the MHD equation, including the bulk viscosity  $\zeta_l$.
The solutions for the magnetoacoustic and sound waves are derived in
Appendix \ref{sec_dispersion}.

To set this problem up, we use a cubic domain, with side length 1.0, is
initialized with density $\rho=1$,
pressure $P=1/\gamma$, and adiabatic index $\gamma=5/3$,  
in which the waves propagate in the $x$-direction.
For the magnetoacoustic wave the magnetic field is $\mathbf{B} = \sqrt{3}\hat{\mathbf{z}}$,
 for the Alfv\'{e}n wave it is $\mathbf{B} = 1.0\hat{\mathbf{x}}$, 
and for the sound wave the magnetic field is set to zero.
The initial condition is generated from a relaxed particle
distribution so that no particle deletion or addition initially occur.  For
runs with resolution higher than the lowest value, the same set of particles used in the
lowest resolution test is scaled and tiled to fill the computational volume.
The wave is propagated one wavelength, and then the $L_1$-norm error is summed
across all fields.  
We analyze the convergence of the relative $L_1$ error, dividing the absolute error
by the amplitude given by the linear analysis  in Appendix~\ref{sec_dispersion},
as for the waves with a compressive nature (sound and magnetoacoustic), the analytical 
solution  varies with resolution.
All the waves are observed to converge with at least second
order accuracy, as shown in Figure \ref{figlinwaveerr}.

\subsection{Circularly Polarized Alfv\'{e}n Wave}
\label{sec_circwaves}

\begin{figure}
\plotone{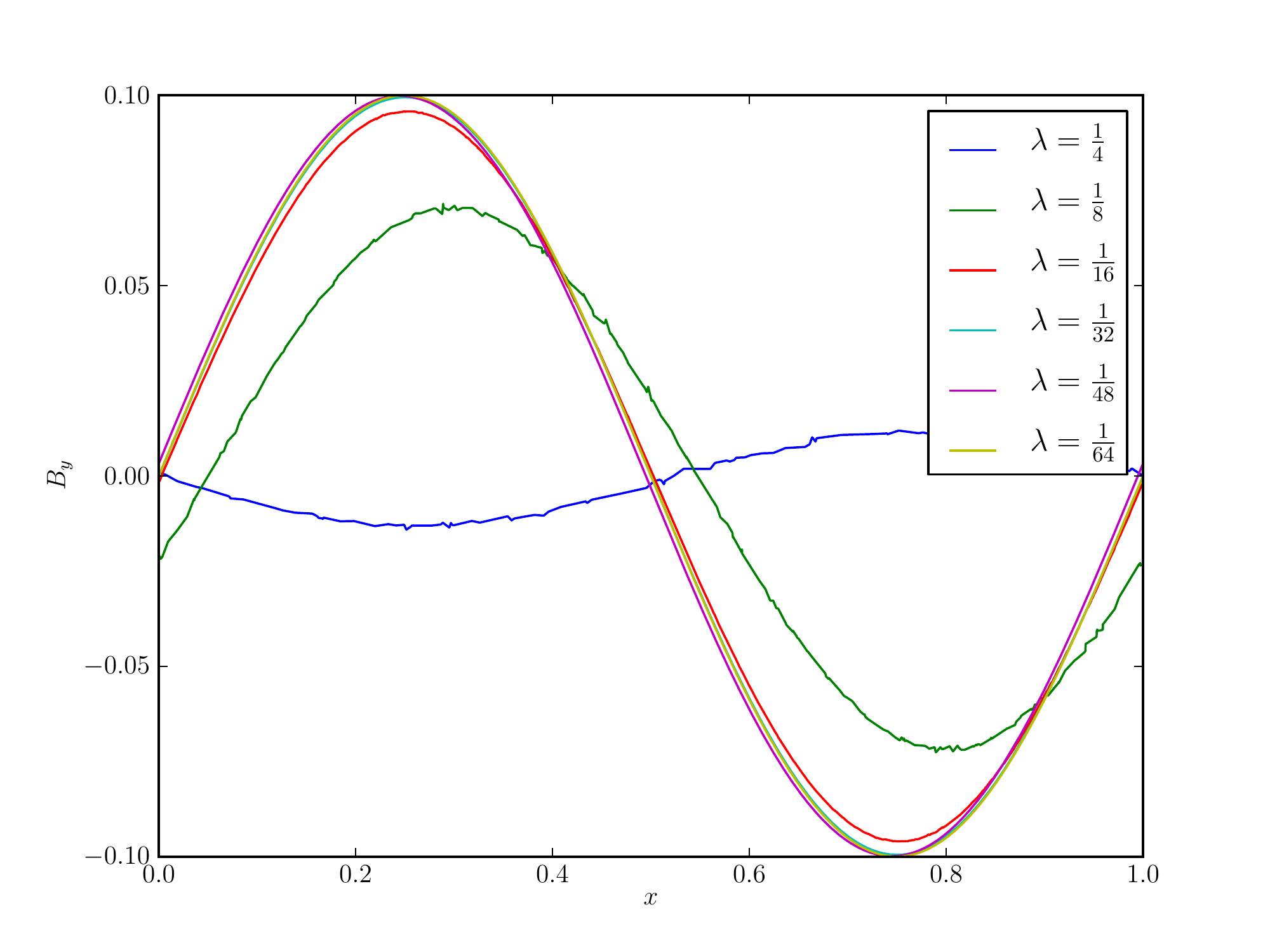}
\caption{Circularly polarized Alfv\'{e}n waves after five box
  crossings with resolutions given by the legend. Note the rapid
  convergence of the higher resolution models.}
\label{figcircalfvensols}
\end{figure}

\begin{figure}
\plotone{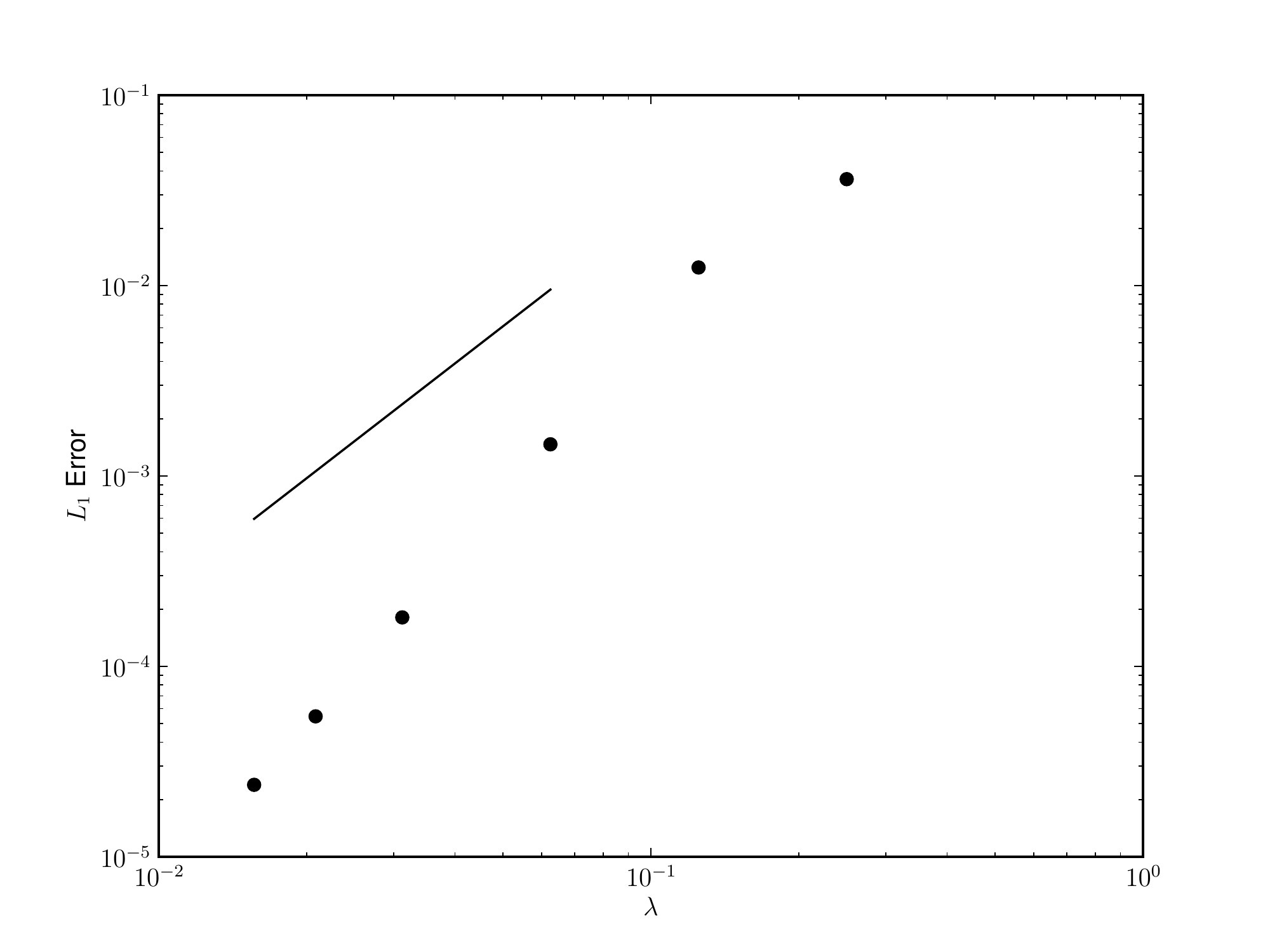}
\caption{ $L_1$-norm errors for circularly polarized Alfv\'{e}n waves
  run with resolution $\lambda$
  after five box crossings. The plotted reference slope is 2.} 
\label{figcircalfvenerr}
\end{figure}

For our next test of MHD, we run finite-amplitude, circularly
polarized, Alfv\'{e}n plane waves using the parameters from
\cite{2000JCoPh.161..605T} in a cubic domain with periodic boundary
conditions, with the propagation direction parallel to the x-axis.
This is equivalent to $\alpha=0$ in the formalism of
\citet{2000JCoPh.161..605T}.  Specifically, the initial conditions are
$\rho=1.0$, $V_x=0.0$, $B_x = 1.0$, $V_y=0.1\sin(2\pi x)=B_y$,
$V_z=0.1\cos(2\pi x)=B_z$, $P=0.1$, and $\gamma=5/3$.  The wave is
propagated five wavelengths, returning to its original position.  The
solutions shown in Figure \ref{figcircalfvensols} show that at low
resolution, the wave is strongly damped, while as resolution increases
the wave rapidly converges.  Figure \ref{figcircalfvenerr} shows the
$L_1$ norm errors, which converge faster than second order.   

\subsection{Shock Tubes}
\label{sec_shocks}

\begin{table*}
\begin{center}
\caption{Shock Tube Left and Right States}
\begin{tabular}{lllllllll}
\hline\hline
Test & $\rho$ & $V_{x}$ &  $V_{y}$ & $V_{z}$ & $P$ & $B_x$ & $B_{y}$ & $B_{z}$ \\
\hline
Sod Left    &  1       &  0    &  0     &  0    & 1     & 0               & 0                 & 0 \\
Sod Right   &  0.125   &  0    &  0     &  0    & 0.1   & 0               & 0                 & 0 \\
Brio-Wu Left & 1       &  0    &  0     &  0    & 1     & 0               & 1                 & 0 \\
Brio-Wu Right& 0.125   &  0    &  0     &  0    & 0.1   & 0               & -1                & 0 \\
RJ2a Left   & 1.08     &  1.2  &  0.01  &  0.5  & 0.95  & $2/\sqrt{4\pi}$ & $3.6/\sqrt{4\pi}$ & $2/\sqrt{4\pi}$ \\
RJ2a Right  & 1        &  0    &  0     &  0    & 1     & $2/\sqrt{4\pi}$ & $4/\sqrt{4\pi}$   & $4/\sqrt{4\pi}$\\
RJ4d Left   & 1        &  0    &  0     &  0    & 1     & 0.7             & 0                 & 0   \\
RJ4d Right  & 0.3      &  0    &  0     &  1    & 0.2   & 0.7             & 0.7               & 1   \\
F6   Left   & 0.5      &  0    &  2     &  0    & 10    & 2               & 2.5               & 0   \\
F6   Right  & 0.1      &  -10  &  0     &  0    & 0.1   & 2               & 2.5               & 0   \\
RJ1a Left   & 1        &  10   &  0     &  0    & 20    & $5/\sqrt{4\pi}$ & $5/\sqrt{4\pi}$   & 0   \\
RJ1a Right  & 1        &  -10  &  0     &  0    & 1     & $5/\sqrt{4\pi}$ & $5/\sqrt{4\pi}$   & 0   \\
\hline
\end{tabular}
\label{tabshocktubes}
\end{center}
\end{table*}

\begin{figure*}
\plotone{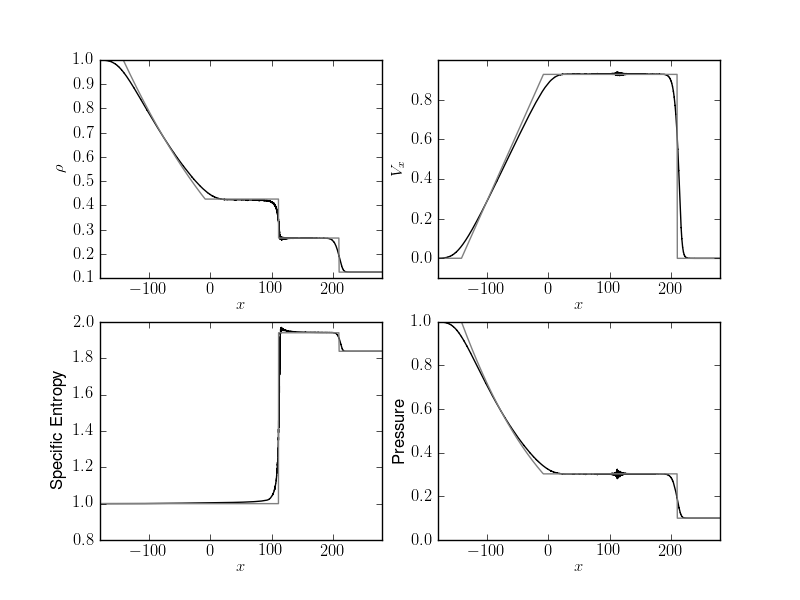}
\caption{\citet{1978JCoPh..27....1S}  shock tube, run with mass-based refinement, equivalent to
a Lagrangian method such as SPH. The $x$-axis is denoted in units of $\lambda$ of the initial left state density. 
Phurbas result in black, reference solution in gray.}
\label{figsodrho}
\end{figure*}

\begin{figure*}
\plotone{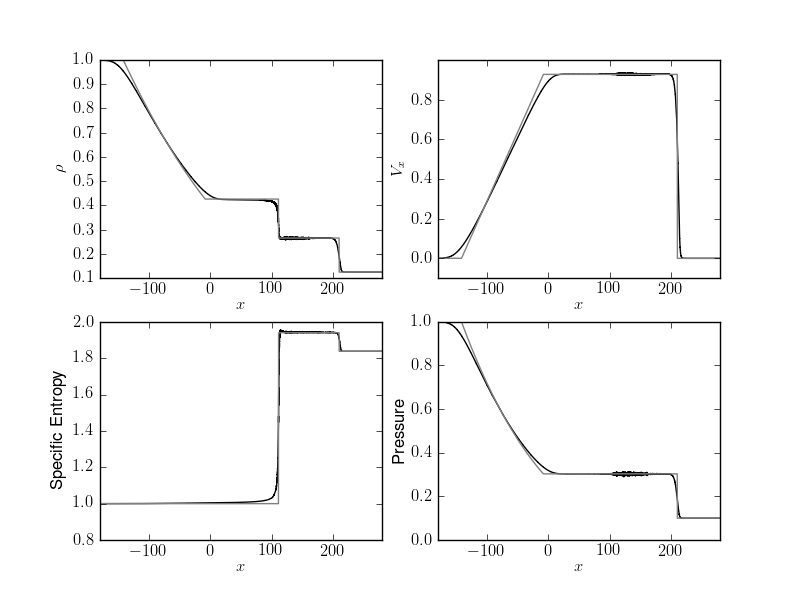}
\caption{\citet{1978JCoPh..27....1S}  shock tube run with constant spatial resolution.
Phurbas result shown in black, reference solution in gray.}
\label{figsod}
\end{figure*}

To test performance on supersonic and super-Alfv\'enic problems, we set
up several shock tube problems.  We used the parameters given in
Table~\ref{tabshocktubes} on long rectangular domains with
periodic boundaries.  
A spatially constant resolution criterion $\lambda$ allows comparison to grid based codes,
while an appropriately density dependent resolution criterion allows
comparison to other Lagrangian methods.
For the constant $\lambda$ tests we denote the distance along the long axis
of the domain in units of $\lambda$.

\subsubsection{Sod Shock Tube}
We first run the classic \citet{1978JCoPh..27....1S} shock tube, 
in gas with $\gamma = 7/5$ 
on a domain of size $64 \times 1 \times 1$
units.  A smooth transition between the left and right states with a
width of $0.3$ units is described with a fifth-order spline.  In the first version 
of this test we use a refinement criteria $\lambda = 0.125\rho^{1/3}$ that
yields a constant mass per resolution element (that is, per region
with volume of $4/3 \pi \lambda^3$).  This is in effect the resolution 
criterion used by SPH.  In Figure~\ref{figsodrho}
the solution at time $t=13.8$ is shown.
For this test the x axis on the plot is denoted in units of $\lambda$ in
the left initial state.  
Unlike SPH,  Phurbas
supports more general refinement criteria. As a simple example, we ran the same Sod test
problem with spatially constant resolution $\lambda=0.125$. Figure
\ref{figsod} shows the result at time $t=15$.
The result is generally very similar to the result with mass-based refinement, 
with the main change being that the shock
is thinner, as the local resolution is higher in the constant-refinement case.
In either case, the shock speed is reasonably well reproduced.

\subsubsection{MHD Shock Tubes}

We next perform a suite of MHD shock tube tests, selecting from the
large set of standard MHD Riemann problems used in the literature.  To
tabulate reference solutions, we have used Athena
\citep{2008ApJS..178..137S} at high resolution 
in 1D.  The first test we show is the classical Brio-Wu
\citeyear{1988JCoPh..75..400B} shock tube test, which is a standard
problem, though not particularly stringent.  The test shown in
Figure~2a of \citet[][denoted RJ2a]{1995ApJ...442..228R} provides a
more complete test of the appearance of the various possible MHD
discontinuities.  \citet{1995ApJ...442..228R} in their Figure~4d show
a test (denoted RJ4d) of other discontinuities.  The problem described
by \citet{2002ApJ...577L.123F} in his Figure~6 (denoted F6) is
specifically used to demonstrate shock errors in
non-locally-conservative methods. Finally, the test shown in Figure~6 of
\citet{1994JCoPh.111..354D}, as well as in Figure 1a of
\citet[][denoted RJ1a]{1995ApJ...442..228R} is commonly used as a stringent test of
$\divb$ errors in shocks, and in the case of Phurbas demonstrates the
effects of local non-conservation errors associated with strong MHD
shocks.  Note that with the exception of the Brio-Wu test, all these
MHD shock tube tests use an adiabatic gas with $\gamma=5/3$.

\begin{figure*}
\plotone{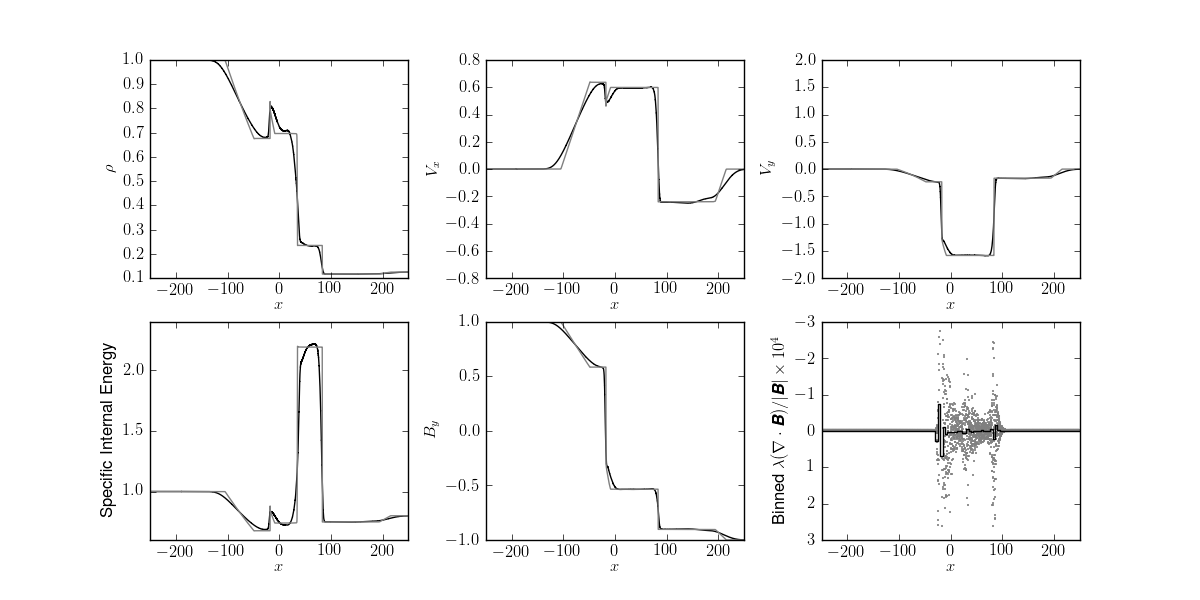}
\caption{Brio-Wu \citeyear{1988JCoPh..75..400B} shock tube solution
  with Phurbas solution in black, and high-resolution Athena solution in gray. 
Lower right panel: One in ten unbinned values of $\divb$ as grey points, binned values as 
black steps. }
\label{figbriowupara}
\end{figure*}

\paragraph{Brio-Wu} The Brio-Wu \citeyear{1988JCoPh..75..400B} shock tube (see 
Table~\ref{tabshocktubes}) is an MHD analog 
to the Sod shock tube problem.  We set up this test
in gas with $\gamma = 2$, on a fully periodic domain of
$128\times 1\times 1$ units, with constant resolution $\lambda =
0.125$.   A smooth transition between the left and right
states with a width of $3$ units was produced with a cosine function.
The width of the transition region was chosen to avoid excessive
start-up transients. 

As the problem was run in two mirror images in a periodic volume, only 
 half of the periodic volume used is shown 
in Figure \ref{figbriowupara}, with the $x$-axis labeled in
$\lambda$ units, at time $t=7.3$.  
The solution captures the fundamental features of
the problem, including, from left to right, the rarefaction fan, the compound
wave, the contact discontinuity, the slow shock, and the fast
rarefaction wave.    

The final panel of Figure~\ref{figbriowupara}
shows a measure of the effect of $\divb$.  Here the quantity
$\lambda(\divb)/|\sB|$ is the fractional magnetic field error
on the scale of $\lambda$.  Grey points in this panel show the raw
values of this quantity, with maximum magnitude  $10^{-3}$.
However, the scatter is very symmetric, indicating that most of it can
be attributed to the truncation error in the approximation of $\divb$
itself.  To extract the coherent skew from zero, we plot the
data binned in bins with width $\lambda$ as the black step curve,
demonstrating that the normalized $\divb$ errors are
less than $10^{-4}$.

The Athena solution that we compare our result to, as well as the
usually accepted numerical solutions to the Brio-Wu problem, show a
compound wave structure, seen at $x\approx -25$ in
Figure~\ref{figbriowupara}.  This structure
should not formally exist \citep{2001JPlPh..65...29F}, but most
multidimensional numerical methods, including Phurbas, show it as part
of the solution.

\begin{figure*}
\plotone{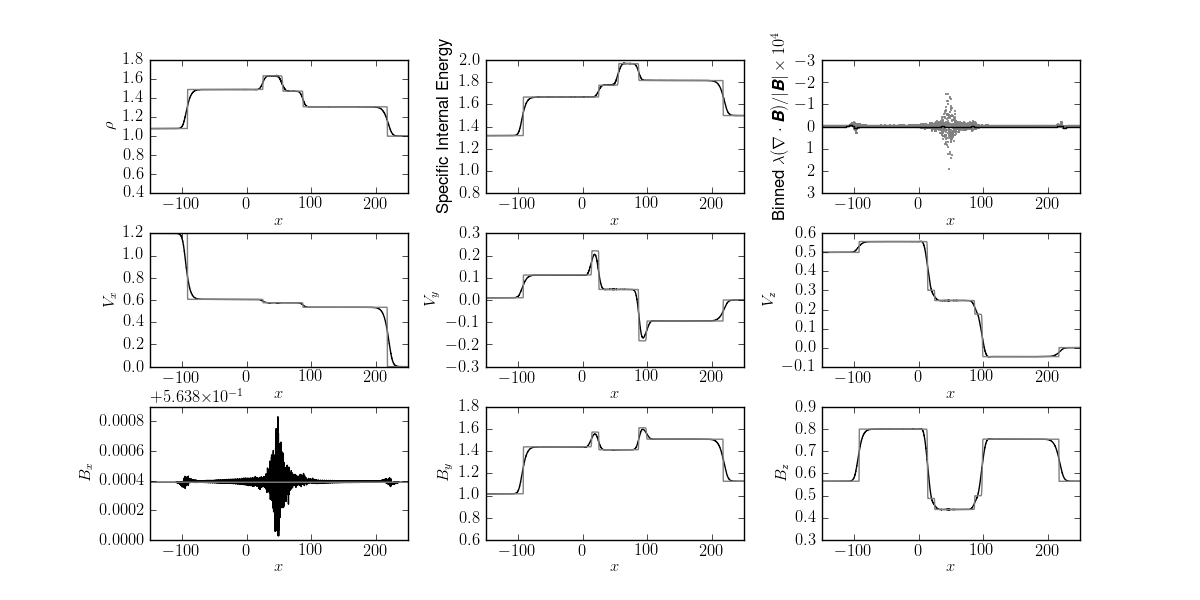}
\caption{
\cite{1995ApJ...442..228R} test from their Figure 2a (model RJ2a). 
High resolution Athena solutions are shown in 
gray, while the normalized $\nabla \cdot B$ is shown as
gray points, and the binned values are shown in black steps. 
One in ten particles is plotted in the $\divb$ scatter.}
\label{figrj2a}
\end{figure*}

\paragraph{RJ2a} The test shown in Figure 2a of \cite{1995ApJ...442..228R}  (also see
\cite{1994JCoPh.111..354D} Table 3a) 
is shown in Figure~\ref{figrj2a} at time $t=12$. Initial
conditions for this test are listed in Table~\ref{tabshocktubes} as RJ2a.
This test is otherwise set up in the same manner and on the same
domain as the Brio-Wu test.  
A high resolution solution was computed
with Athena for comparison.  As \citet{2008ApJS..178..137S} point out,
this test is particularly interesting because it requires modeling a
fast magnetosonic shock and a rotational discontinuity in each
direction of propagation, as well as a contact discontinuity in the
center.  No visible ringing is seen at the shocks.
The largest coherent $\divb$ errors are also seen near the fast shocks, 
but the largest scatter in particle $\divb$ values
 is located at the contact discontinuity.

\begin{figure*}
\plotone{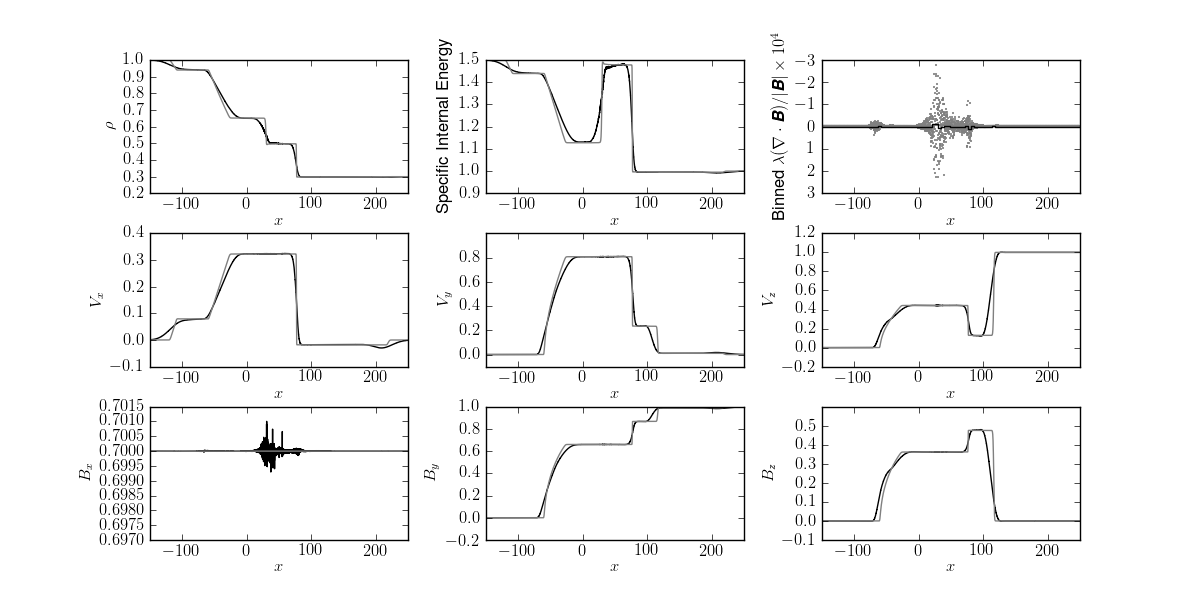}
\caption{\cite{1995ApJ...442..228R} test from their Figure 4d (model
RJ4d). High resolution Athena solution shown in
gray, while the normalized $\divb$ is shown as
gray points, and the binned values are shown in black steps. 
One in ten particles plotted in $\divb$ scatter.}
\label{figrj4d}
\end{figure*}

\paragraph{RJ4d} The test shown in Figure 4d of \cite{1995ApJ...442..228R} is shown in
Figure~\ref{figrj4d} at time $t=11.5$. 
This test was run with the same
computational domain and resolution as the 
previous test but with the left and
right states listed in Table~\ref{tabshocktubes} as test RJ4d.  
A small overshoot can be seen on the leftmost rarefaction wave. 
This is a result of the bulk viscosity of the scheme.

\begin{figure*}
\plotone{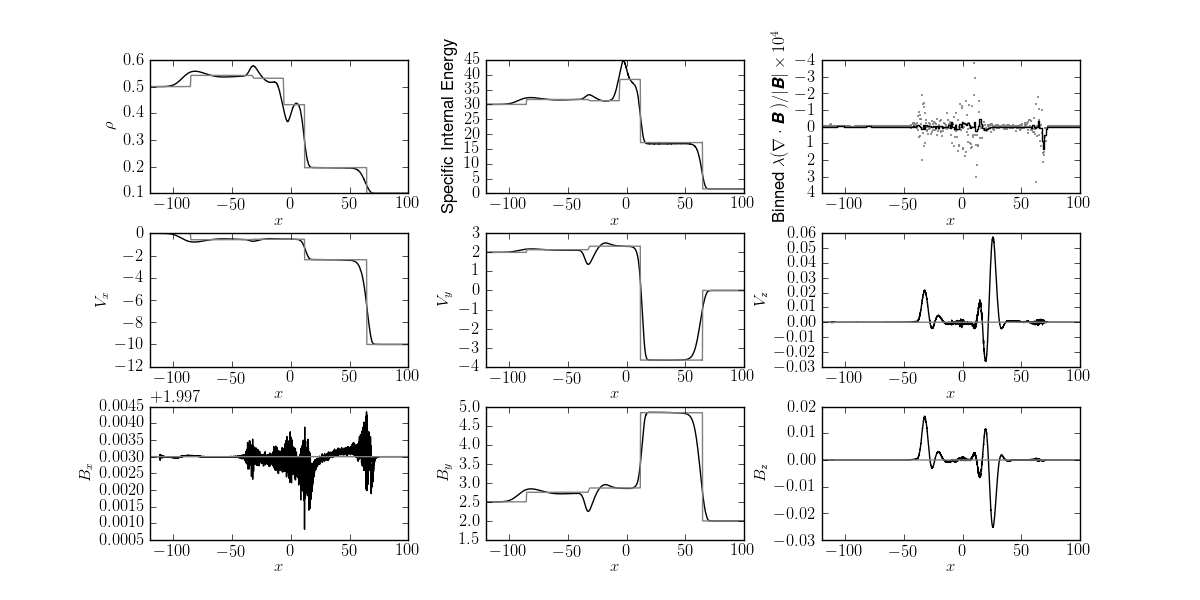}
\caption{\cite{2002ApJ...577L.123F} test from their Figure~6 (our test
F6), which was used to demonstrate an error in the ZEUS algorithm. 
High resolution Athena solution shown in 
gray, while the normalized $\divb$ is shown as
gray points, and the binned values are shown in black steps. 
One in ten particles is plotted in $\divb$ scatter.}
\label{figfalle2002}
\end{figure*}

\paragraph{F6} The test shown by \cite{2002ApJ...577L.123F} in his Figure 6 
is shown in Figure \ref{figfalle2002} at time $t=2.9$.  This test was
used to show an error in the {\sc ZEUS} \citep{1992ApJS...80..791S} solution.  
The initial states are listed in
Table \ref{tabshocktubes} as test F6 and the problem was run in a
domain of $64
\times 1 \times 1$ units
with $\lambda=0.25$.
Fixed boundaries were used in the $x$ direction and periodic
boundaries in the $y$ and $z$ directions.  
Compared to the nonconservative {\sc
ZEUS} result shown in \cite{2002ApJ...577L.123F} the slow shock is captured
more accurately.  
At this resolution, the slow shock and the contact discontinuity 
directly behind it have not yet separated, and the bulk viscosity of the scheme
is causing an undershoot in density at the foot of the contact discontinuity.
The left-going features also show overshoots in density and specific internal 
energy that are initial transients caused by the bulk viscosity.
These effects are not however due to significant nonconservation errors.

\begin{figure*}
\plotone{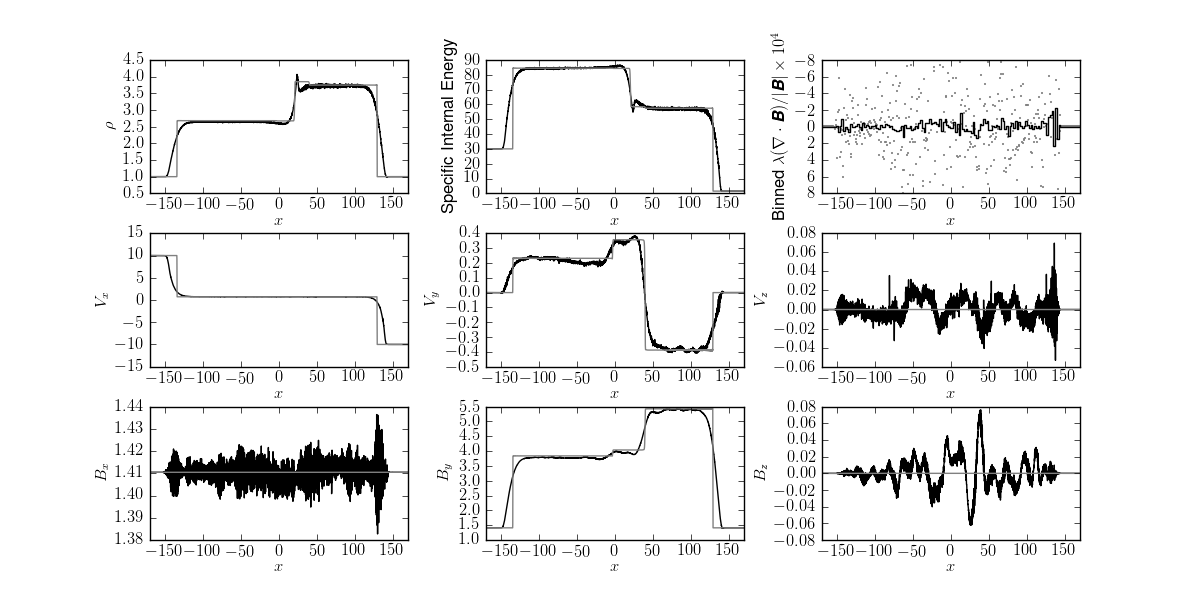}
\caption{
Riemann problem test shown by \cite{1995ApJ...442..228R} in their Figure 1a, run without 
elliptic projection $\divb$ correction. 
High resolution Athena solution shown in  
gray, while the normalized $\divb$ is shown as
gray points, and the binned values are shown in black steps. 
One in ten particles is plotted in the $\divb$ scatter.}
\label{figtothpara}
\end{figure*}

\paragraph{RJ1a} The test shown by \cite{1994JCoPh.111..354D} in their Figure 6, is shown in
Figure~\ref{figtothpara} at time $t=3.5$.  This test also appears in
\cite{1995ApJ...442..228R} in their Figure~1a, in Table~6 of
\cite{2000JCoPh.161..605T}, and in \cite{2010JCoPh.229.5896M}. In the
latter two it is used as a demonstration of $\divb$ errors.
We show this problem here as a test of the technique used in Phurbas
to handle $\divb$, though we expect that the strength of the shocks in
this problem lies outside of the intended use regime of Phurbas.
A computational volume $64\times 1\times 1$ units was used, with fixed
value boundaries in the $x$ direction, and periodic boundaries in the
$y$ and $z$ directions with left and right states listed in 
Table~\ref{tabshocktubes} as test RJ1a.
As \cite{1994JCoPh.111..354D} show, the
solution consists of a right going fast shock with Mach number 6.54, a
left going fast shock with Mach number 2.54, a slow shock, a slow
rarefaction, and a contact discontinuity.  Phurbas shows a $\nabla \cdot \mathbf{B}$
error of a few parts in $10^4$ in the region lying between the two fast shocks.

\subsection{Kelvin-Helmholtz Instability}
\label{sec_khi}

As an example of multidimensional smooth flow with bulk motions, we demonstrate
the performance of Phurbas on three-dimensional Kelvin-Helmholtz instability.  Recently
Kelvin-Helmholtz instability has attracted significant discussion following the
demonstration by \citet{2007MNRAS.380..963A} that some SPH formulations 
fail to show the growth of the instability in a particular test problem.  A
number of works have discussed aspects of Kelvin-Helmholtz instability
in numerical methods
used in astrophysics, particularly in Lagrangian schemes
\citep{2008JCoPh.22710040P,2008MNRAS.387..427W,2010MNRAS.403.1165C,2010MNRAS.406.2289H,2010MNRAS.401..791S,2010MNRAS.405.1513R,2010MNRAS.407.1933J,2010MNRAS.408...71V,2010ARA&A..48..391S,2010MNRAS.401.2463R,2011arXiv1109.2218S}.

In the terminology of \cite{2010MNRAS.401.2463R} and
\cite{2010ARA&A..48..391S} we use a smoothed interface initial
condition.  However, unlike \cite{2010ARA&A..48..391S} we choose a
smoothing of the interface such that a closed-form algebraic
expression can be computed for the first order, linear, perturbation
theory prediction for the growth rate in an incompressible flow on an
infinite domain.  \cite{2010PhPl...17d2103W} have derived such an
analytic treatment, providing an exact analytic benchmark for the
first time.  From that work, we select an initial condition with an
exponential smoothing.  The initial condition for the Kelvin-Helmholtz
test is as follows in coordinates where $x$ is parallel to the
direction of flow, and $z$ is in the direction of slab symmetry.  Density
is given by
\begin{equation}
\rho = \left\{ \begin{array}{ll}
\rho_1 - \rho_m e^\frac{y-1}{L} &\mbox{ if }     1>y>0\\
 \rho_2 +\rho_m e^\frac{-y+1}{L} &\mbox{ if } 2>y>1\\
 \rho_2 +\rho_m e^\frac{-(3-y)}{L} &\mbox{ if } 3>y>2\\
 \rho_1 -\rho_m e^\frac{-(y-3)}{L} &\mbox{ if } 4>y>3,
\end{array}
\right.
\end{equation}
where $\rho_1 = 1.0$, $\rho_2 = 1.1$, $\rho_m = (1/2)(\rho_1 - \rho_2)$, and
$L=0.025$.
The velocity field is given by a similar smoothed profile with a
perturbation in the $x$ direction 
\begin{equation}
V_x = \left\{ \begin{array}{ll}
U_1 - U_m e^\frac{y-1}{L} &\mbox{ if } 1>y>0\\
U_2 + U_m e^\frac{-y+1}{L} &\mbox{ if } 2>y>1\\
U_2 + U_m e^\frac{-(3-y)}{L} &\mbox{ if } 3>y>2\\
U_1 - U_m e^\frac{-(y-3)}{L} &\mbox{ if } 4>y>3,
\end{array}
\right.
\end{equation}
where $U_1 = 0.5$, $U_2 = -0.5$, $U_m = (1/2)(U_1 - U_2)$, and
a perturbation in the $y$ direction
\begin{equation}
V_y = \delta v  \sin(4\pi x) \left\{ \begin{array}{ll}
\exp(4\pi (y-1))  &\mbox{ if } 1>y>0\\
\exp(4\pi (-y+1)) &\mbox{ if } 2>y>1\\
\exp(4\pi (-(3-y)) &\mbox{ if } 3>y>2\\
\exp(4\pi (-(y-3)) &\mbox{ if } 4>y>3,
\end{array}
\right.
\end{equation}
where $\delta v = 0.01$.
Pressure is set to $2.5$, and $\gamma=5/3$.

To extract the growth of the velocity perturbation, we use a discrete
convolution over the particles.
We use this technique, as opposed to a discrete Fourier transform
on gridded values, as the discrete 
convolution maps more 
directly to the meshless Phurbas discretization.
We define the magnitude of the unstable $y$-velocity mode
\begin{align}
M = 2\left[\left(\frac{\sum {s_i}}{\sum{d_i}}\right)^2+\left(\frac{\sum{c_i}}{\sum{d_i}}\right)^2\right]^\frac{1}{2},
\end{align}
where all the sums run over $N$ particles, and
\begin{align}
s_i &=\left\{\begin{array}{l l} V_y\sin(4\pi x)\exp(-4\pi|y-1|) & \mbox{ if } y<2\\
 V_y\sin(4\pi x)\exp(-4\pi|(4-y)-1|) & \mbox{ if }y>2,\\
\end{array}\right. \\
c_i &=\left\{\begin{array}{l l} V_y\cos(4\pi x)\exp(-4\pi|y-1|) & \mbox{ if } y<2\\
 V_y\cos(4\pi x)\exp(-4\pi|(4-y)-1|) & \mbox{ if }y>2,\\
\end{array}\right. \\
d_i &=\left\{\begin{array}{l l} 
\exp(-8\pi|y-1|) & \mbox{ if } y<2\\
 \exp(-8\pi|(4-y)-1|) & \mbox{ if }y>2.\\
\end{array}\right. \\
\end{align}
The domain is $4\times1$ units with thickness $1/32$, $1/64$, or
$1/128$, simulated with uniform target resolutions $\lambda=1/64$,
$1/128$, and $1/256$.  

\begin{figure}
\epsscale{0.5}
\plotone{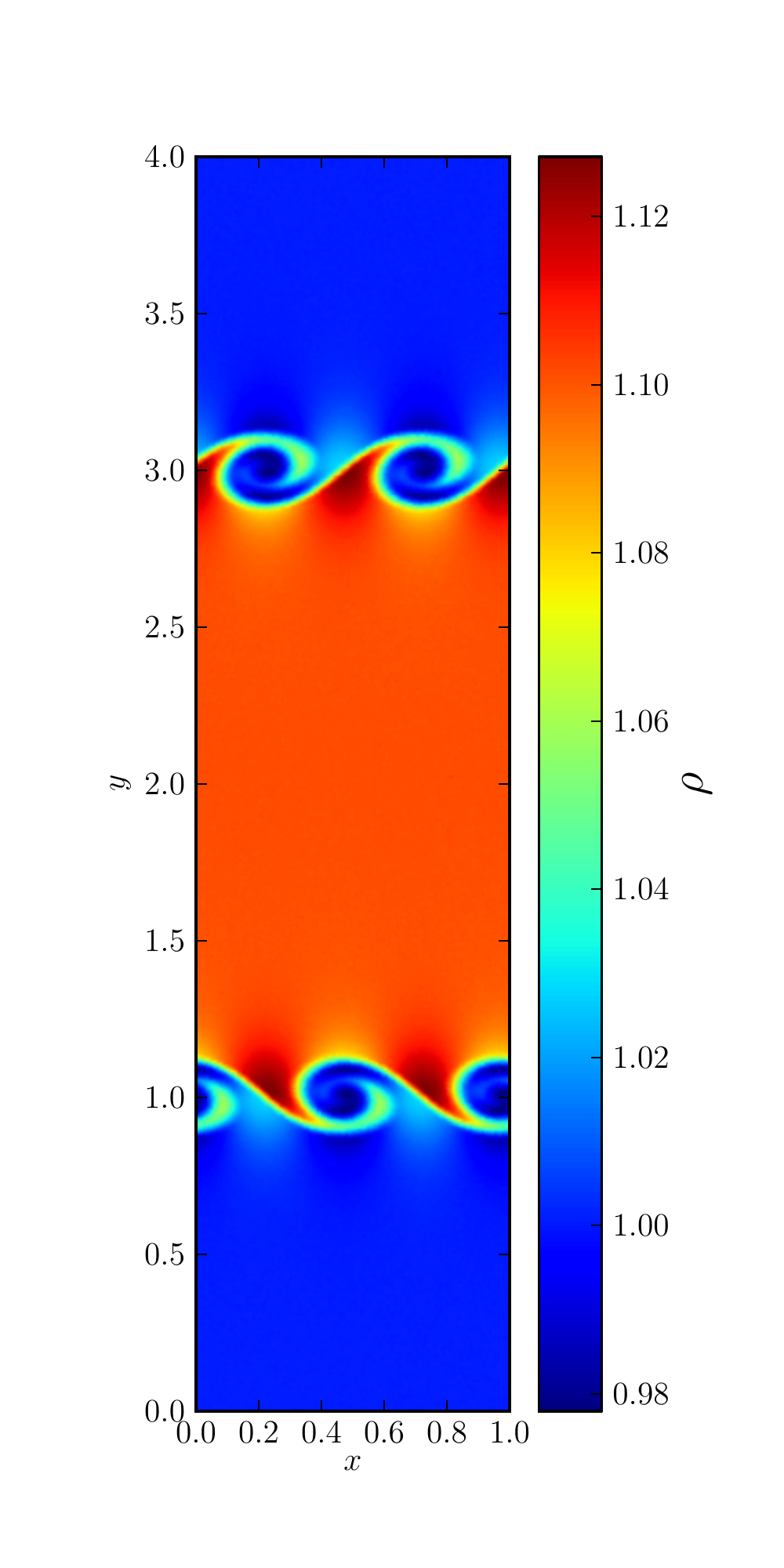}
\epsscale{1.0}
\caption{Kelvin-Helmholtz instability test presented at the highest
  resolution used here ($\lambda = 1/256$). A density slice is shown at t=2.52.}
\label{figkhslice}
\end{figure}

\begin{figure}
\plottwo{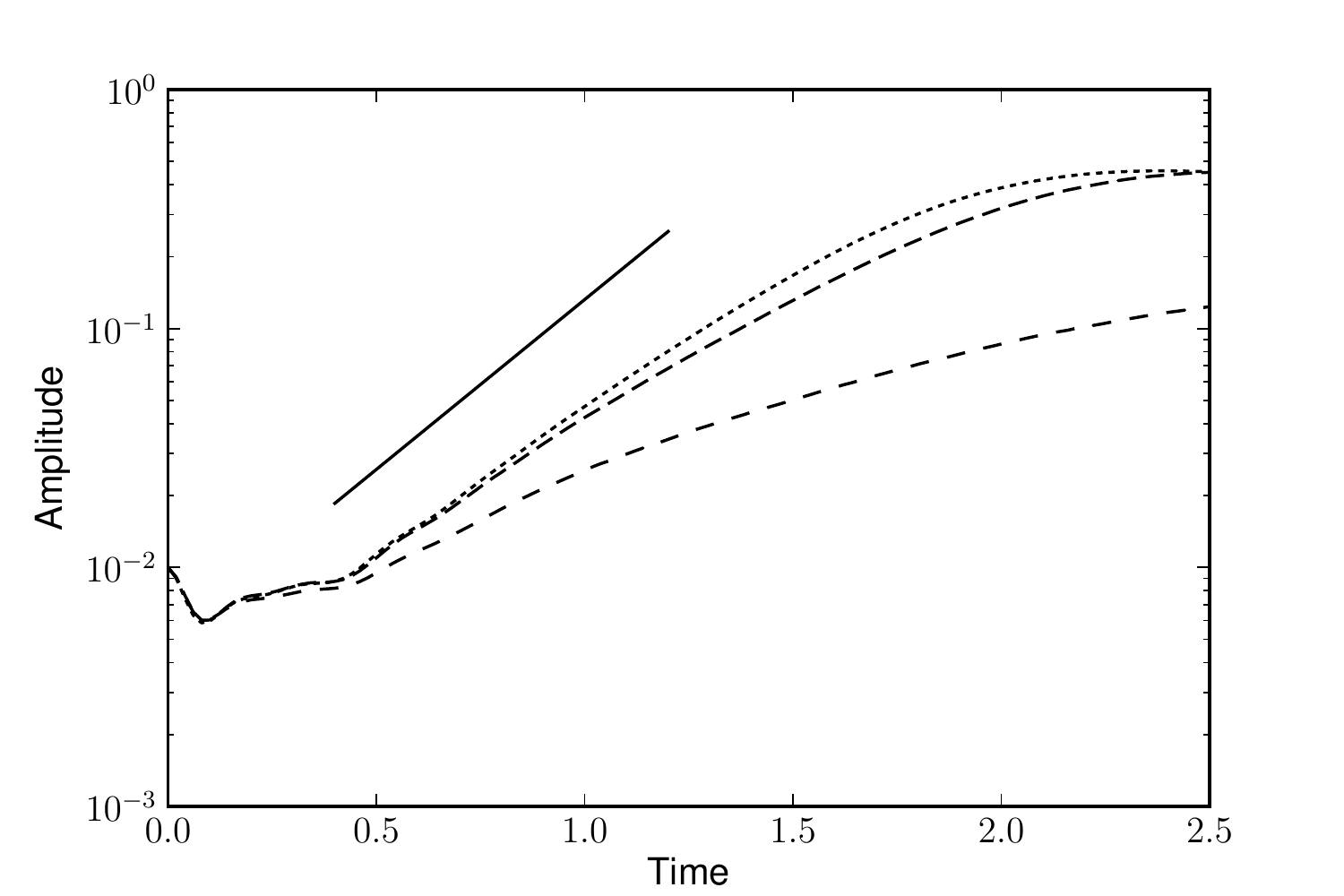}{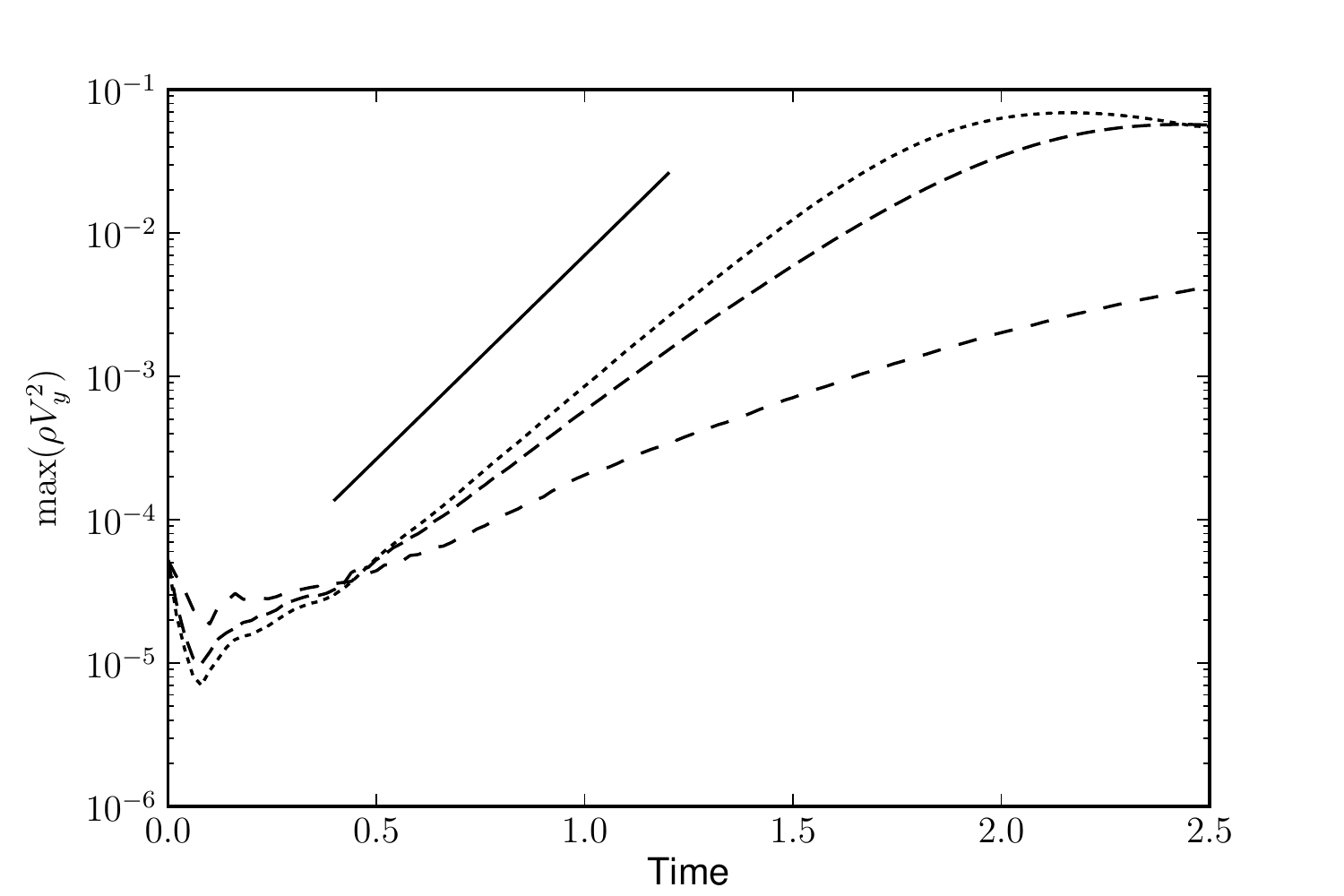}
\caption{Left: Kelvin-Helmholtz instability mode growth for models with
$\lambda = 1/64$ (dashes), $1/128$ (tight-spaced dashes), and $1/256$
units (dotted), compared to 
  the predicted growth rate (solid) from the linear, incompressible
  theory derived by \cite{2010PhPl...17d2103W}. 
Right: Maximum $y$-direction kinetic energy for the same runs, again
compared to the predicted growth rate from linear, incompressible theory.}
\label{figkhgrowth}
\end{figure}

In Figure \ref{figkhslice} the developed state
of the highest resolution run is shown, while Figure~\ref{figkhgrowth} shows
the growth of the unstable $y$-direction velocity mode and the maximum
$y$-direction kinetic energy.  The linear perturbation theory predictions
from \cite{2010PhPl...17d2103W} for the growth rate of the mode
amplitude and the maximum $y$-direction kinetic energy are also
plotted as a guide, but due to the periodic domain, compressibility,
and finite perturbation magnitude we cannot expect the result to
exactly follow this curve.  However, clear convergence toward the
linear theory can be seen as resolution increases.  A more
sophisticated, multiple code verification study, circumventing the
limitations of the incompressible theory comparison for a similar
problem, will be presented in a separate publication.

\subsection{Magnetorotational Instability in a Cylinder} \label{sec_flocktest}

The magnetorotational instability \citep[MRI;][]{1991ApJ...376..214B} is thought
to be an important mechanism for driving turbulence in protoplanetary disks
\citep{1998RvMP...70....1B}.  We have performed a test similar to one done by
\cite{2010A&A...516A..26F} to examine the growth rate of MRI during
its linear phase.  
For this test, we 
remove the background Keplerian shear flow from the velocity field, and 
evolve only the perturbations to the initial steady
state velocity. 
That is, the velocity field we evolve and fit in this test is $\sV' =
\sV - \Omega r\hat{\mathsf{\phi}}$ 
where $r$ is the cylindrical radius from the point (0.5,0.5), 
 $\Omega = r^{-1.5}$ is the Keplerian 
angular velocity and $\hat{\mathsf{\phi}}$ is the
unit vector in the azimuthal direction.
Additionally, to ensure that the magnetic field configuration we study is 
consistent between resolutions, we solve only for perturbations from the
initial magnetic field configuration.
Solving only for the perturbations is particularly useful
here as the MRI grows fastest where the orbital timescale is smallest.

The magnetic field configuration used in this test,
an annulus of uniform vertical field, is chosen to artificially
cut off the growth of MRI at a finite radius.
Ensuring that the MRI is cleanly shut off and that the annulus is represented 
in a constant manner allows clearer 
measurements of the convergence of the growth rate in the magnetized annulus.
The domain is an annulus with radius ranging from 0.08
to 0.32 and height 0.0375 units.  The vertical
boundaries are periodic.  The inner and outer radial boundaries are
fixed-value, arranged by preventing time evolution for particles with radius
greater than 0.3 or less than 0.1.  The initial density is 1.0 everywhere. A vertical
magnetic field is imposed with radial profile $B_z(r) = B_0
b_p(r)$, where $B_0=0.0824$ and
\begin{equation} 
b_p(r) = \left\{\frac{1}{2}+\frac{1}{2}\tanh\left(
5-\sin\left[\frac{2\pi(r-0.05)}{0.2}\right] \right)\right\},
\end{equation} 
which gives a magnetized annulus.  The sound speed
in the magnetized annulus was set to $c_s = 0.824$, and the internal energy in the
non-magnetized region adjusted so that the total pressure (thermal plus
magnetic) is constant.  The radial velocity is perturbed with 
\begin{equation}
V_r(z) = 10^{-5}c_s b_p(r)\cos(2\pi z/0.1). 
\end{equation} 
Spatially constant resolutions of $\lambda=1/240$, $\lambda=1/320$, and $\lambda=1/400$
were used corresponding to $\lambda = 1/9 \lambda_\mathrm{MRI}$, 
$\lambda = 1/12 \lambda_\mathrm{MRI}$,
$\lambda = 1/15 \lambda_\mathrm{MRI}$ where $\lambda_\mathrm{MRI}=0.0375$ is the
wavelength of the fastest growing MRI mode at $r=0.17$. 

\begin{figure*}
\plotone{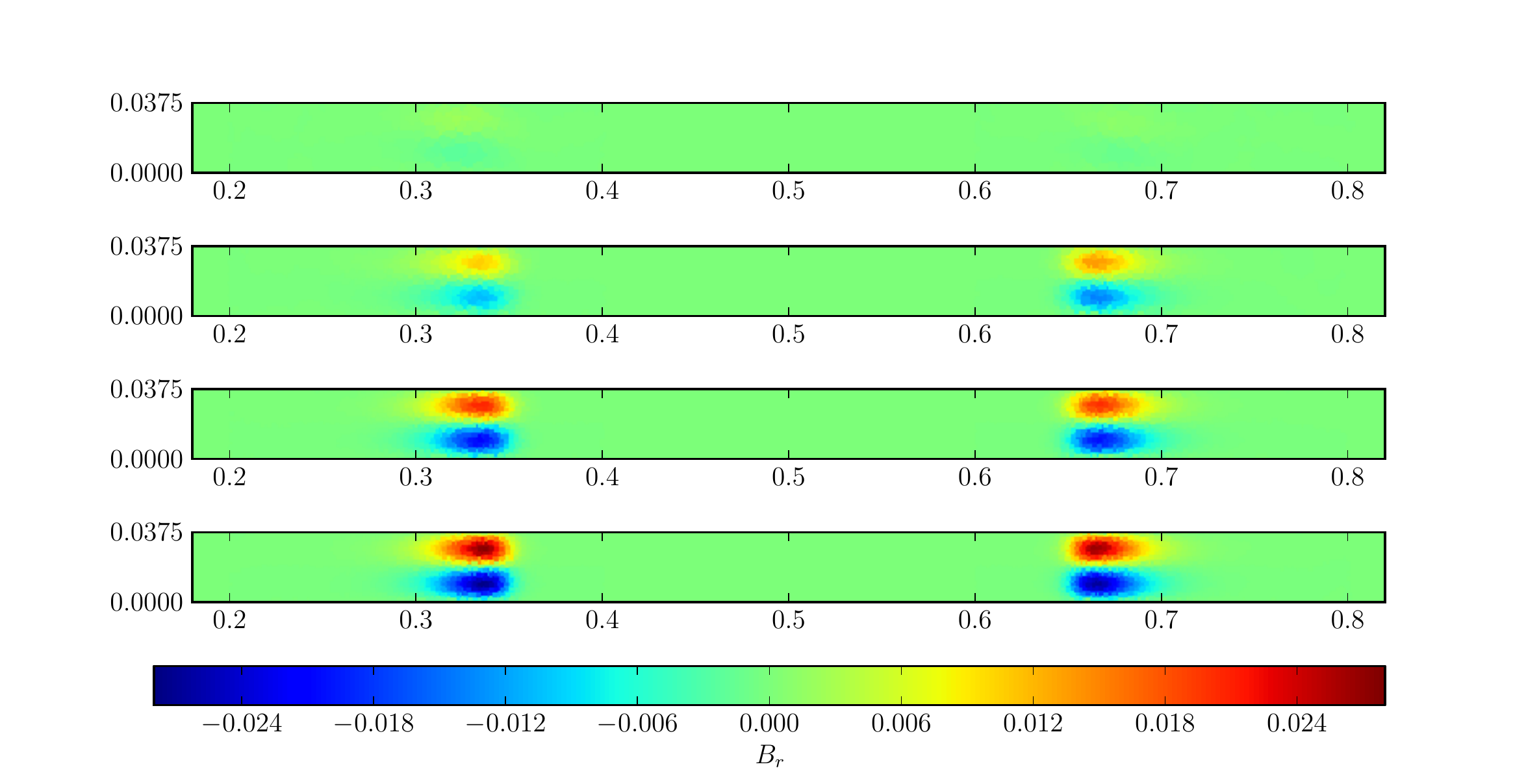}
\caption{Linear phase growth of MRI shown via azimuthal field slices at time 0.94
for resolutions (from top to bottom) of $\lambda = 1/160, 1/240, 1/320$, and $1/400$.}
\label{figflocktestslice}
\end{figure*}

 We then measure the growth of the most
unstable mode at $r=0.17$.  
Figure~\ref{figflocktestslice} shows the radial magnetic field configuration achieved 
at time $0.94$ (or $2.13$ orbits at $r=0.17$) for all three resolutions.
To calculate the mode amplitude $M$, we use a convolution
defined directly on the particles, instead of gridding and Fourier transforming the data.
The motivations are the same as when this procedure was used for the
Kelvin-Helmholtz test.  In this case, 
\begin{equation}
M =
\left[\left(\frac{\sum{s_i}}{\sum{d_i}}\right)^2+\left(\frac{\sum{c_i}}{\sum{d_i}}\right)^2\right]^\frac{1}{2}, \label{eq_KHImode}
\end{equation}
where all sums run over the $N$ points, and
\begin{align}
s_i &= B_{r,i} \sin\left(\frac{2\pi}{0.0375} z_i\right)\exp\left(-\frac{(r_i-r_0)^2}{\sigma^2}\right),\\
c_i &= B_{r,i} \cos\left(\frac{2\pi}{0.0375} z_i\right)\exp\left(-\frac{(r_i-r_0)^2}{\sigma^2}\right),\\
d_i &= \exp\left(-\frac{(r_i-r_0)^2}{\sigma^2}\right).
\end{align}
The chosen width of the convolution $\sigma=2.7\times10^{-6}$ minimizes the radial 
range that influences the measurement, while still giving sufficiently low sampling noise. 

\begin{figure}
\plotone{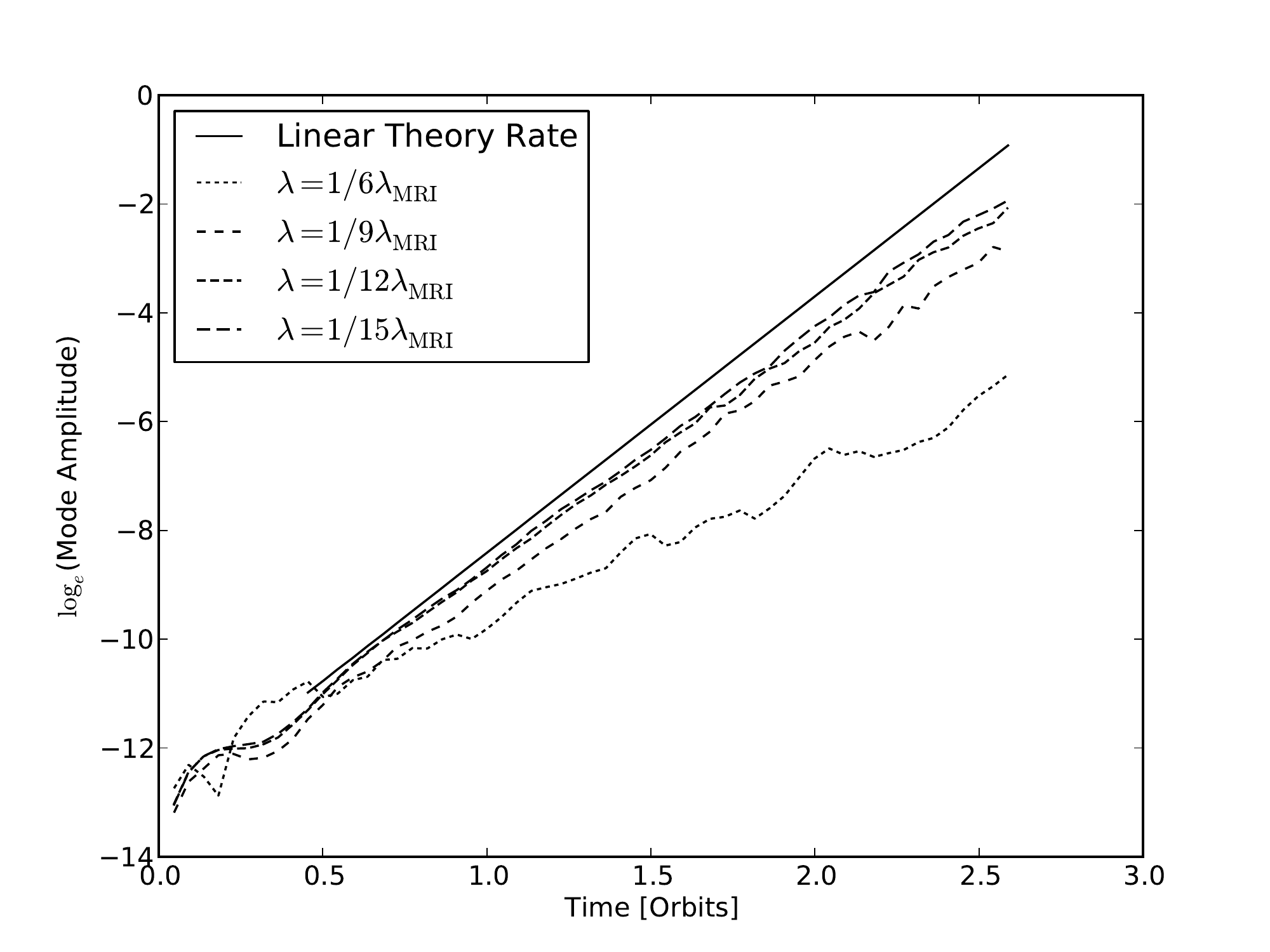}
\caption{Linear phase growth of MRI shown for the mode amplitude given
  by equation~(\ref{eq_KHImode}) at $r=0.17$ for the
  resolutions $\lambda$ given in the legend, compared to the growth
  rate predicted by linear theory.}
\label{figflocktest_18}
\end{figure}

We plot the evolution of the mode amplitude in Figure~\ref{figflocktest_18} together with
the maximum growth rate of $\exp(0.75\Omega)$ predicted by a linear perturbation analysis of
vertical field MRI. The modeled growth rates 
are reasonably consistent with the prediction from the linear analysis
for the fastest growing mode.
Importantly, in the context of the findings of
\cite{2010A&A...516A..26F}, where spuriously high growth rates were
observed, we find growth rates slightly lower
than the theoretical maximum value.


\section{Summary and Discussion}

\label{sec_discussion}

\subsection{Effective Resolving Power}

To determine if Phurbas can be used for practical computations, we
need to establish some guidelines for its relative ability to resolve
particular phenomena. 
This should be done cautiously, as different classes of algorithms have 
different properties in each flow regime. An equivalence or difference between
algorithms in one regime may not hold across different regimes.  
In any case, it is expected that an unstructured mesh or unstructured meshless
method will have a lower effective resolution than a structured mesh method.
This is because a given number of resolving elements can represent the largest possible 
set of wavelengths when they are arranged in a regular manner.
Given these caveats, we can compare the test results that we have
presented here to examples computed with Eulerian, mesh-based schemes.

The first example is the circularly polarized Alfv\'{e}n wave test.
The lowest resolution, three-dimensional, Athena results in a
rectangular domain published in \citet[][Fig. 33]{2008ApJS..178..137S}
correspond to $20$ and $39$ cells per wavelength, computed with third
order spatial reconstruction and HLLD fluxes.  The Phurbas results on
this test at $\lambda = 1/8$ and $\lambda = 1/16$ of a wavelength
appear to roughly equal the accuracy of the $20$ and $39$ cells per
wavelength Athena results in the sense that the final amplitude of the
wave in the Phurbas results is closer to the analytically correct
value even though there are fewer resolution elements used per
wavelength. This result should be interpreted cautiously, as the two
codes have different and nonlinear numerical dissipation.
Nevertheless, this can be interpreted to mean that the Phurbas
effective resolution $\lambda$ is roughly equal to two Athena cells as a
measure of resolution.  
(On average, there should be one particle in each volume of radius
$\lambda$.)  In this sense Phurbas with a third-order polynomial fit
is competitive with a spatially third-order grid code.

A possibly more operationally useful comparison can be drawn from 
the results of the linear phase MRI growth test.
\citet{2010A&A...516A..26F} claim that in the code 
Pluto \citep{2007ApJS..170..228M} with piecewise linear 
reconstructions and an HLLD Riemann solver, 10 cells per wavelength
are required to resolve the growth of MRI.
Our test in section \ref{sec_flocktest} shows Phurbas requires
$\sim 9$ $\lambda$ per MRI wavelength to resolve the growth.
Thus, for this test we can say that one~cell$\ \approx 1 \lambda$. 
The algorithm used by \citet{2010A&A...516A..26F} has a stencil size
of five cells, while
Phurbas can be said to have a stencil size of $2 r_f =4.6\lambda$, so
the same rough proportionality holds between the two algorithms when
expressed in terms of stencil size.

\subsection{Advantages and Disadvantages}

The main advantage of Phurbas is its Lagrangian nature.  Eulerian codes suffer
from numerical dissipation that varies with the speed and direction of the
flow across the grid.  Phurbas's formulation cleanly avoids this behavior.
For systems where the bulk velocity varies as a multiple or large fraction of the wave
speeds, this means Phurbas can capture the flow with more uniform fidelity across
the domain.  Techniques that add an extra advection step to an Eulerian method
\citep[e.g.][]{2000A&AS..141..165M,2009ApJ...697.1269J} can only efficiently handle simple
flow geometries.  Moreover, in Phurbas, the time step is only dependent on
Galilean-invariant quantities.  For flow with bulk velocities greater than the
signal speeds that determine the Courant time step limit, Phurbas has a
significant advantage in the number of time steps that must be
computed to reach a given physical time.

The resolution criterion in Phurbas is spatially continuous, unlike in
adaptive mesh refinement techniques, so there are no abrupt resolution
jumps that can lead to undesirable artifacts.  Also, due to the
Lagrangian and meshless nature of the code, refined regions can be arbitrarily
shaped and follow the flow with minimal creation and deletion of
resolution elements.

As Phurbas formally computes strong solutions to the governing partial
differential equations using a method-of-lines type approach, it is relatively
simple, compared to Godunov methods for example, to switch to an alternate set
of variables or even alternate equations.  The central limitation is only that the
equations should be amenable to the explicit Hermitian time
integration used.  However, changing the time integration scheme would not
fundamentally alter the method.

A moving unstructured mesh or meshless method must win in terms of the
Galilean invariance of the numerical diffusion, adaptivity, and the
time step advantages, since
the quality of the spatial derivatives on an unstructured set of
discretization points is lower than for equivalent fits on
a grid of points. For problems where the Lagrangian and adaptive nature of Phurbas is of 
no significant advantage, a grid-based method such as the Pencil Code
\citep{2002CoPhC.147..471B} remains substantially more efficient.

\subsection{Future Prospects}

Several enhancements to Phurbas can be made, which we briefly mention here.  
\begin{itemize}
\item The
Phurbas discretization is very flexible in how it can incorporate 
governing equations other than ideal MHD.  
Implementing non-ideal MHD is relatively simple with
forward-time-centered-space viscosity and resistivity operators.  
\item Self-gravity
can be implemented using the existing \gadget{} tree algorithm and particle
masses derived from a local Voronoi tessellation.
\item Passive tracer particles and interacting dust
particles can be added in a straightforward manner, using the spatial fitting
and time integration methods used for gas particles.  
\item In the 
moving least squares procedure 
the choice of least
squares error and Cartesian polynomial functions for the fitting procedure 
is not
obviously the ideal choice, and alternate fitting procedures should be
explored. 
These could include those with a basis that can be used to minimize
the variation or oscillation  of the fitted function, and fitting the magnetic
field using a set of divergence-free basis functions \citep{2011MNRAS.413L..76M}.
\item Least-squares minimization itself does not appear to be a unique choice, and less computationally
intensive gradient approximation procedures could be used.
\item Non-Lagrangian particle trajectories can be added, using similar logic to the
steering of Voronoi cell generating centers in \citet{2010MNRAS.401..791S}. 
This could reduce the number of particle additions and deletions, and hence diffusivity
in shear flows. 
\item An improved void check algorithm could reduce the number of redundant
particle creation requests, 
minimizing the potentially expensive step of pruning of the proposed particle addition list.
\item A diffusion with properties similar to a hyperdiffusion would be of great advantage 
if applied to the density and internal energy fields to smooth out the smallest scale structures. 
The challenge of this enhancement is to find an approximation with
sufficiently robust conservation properties.
\end{itemize}
Evidently, there are many possible alterations and extensions that can be made to Phurbas that
will significantly alter both the nature of the scheme and the capabilities of the code.
We believe that Phurbas is not just a new method for MHD,
but one of the first practical examples of a new class of schemes for
mathematical modeling of similar physical problems.

\acknowledgements
Simon C.O. Glover wrote the initial version of the routine to couple
the Phurba MHD module to \gadget{}, which lies at the core of Phurbas.
The tests shown here are of the version of the code incorporating
interpolating fits as advised by the anonymous referee.  As noted in
Paper I, this was a contribution equivalent to co-authorship.
We are indebted to Volker Springel for making the source code of
\gadget{} publicly available 
\citep{2005MNRAS.364.1105S}.
We also thank the authors of Athena for making it publicly available
\citep{2008ApJS..178..137S}.
M.-M.M.L.~and C.P.M.~acknowledge hospitality from the
Max-Planck-Institut f\"ur Astronomie, and M.-M.M.L~additionally
acknowledges hospitality of the Institut f\"ur Theoretische
Astrophysik der Uni.\ Heidelberg.  This work has been supported by
National Science Foundation CDI grant AST-0835734 and allocation
TG-MCA99S024, originally from the Teragrid, and now from the Extreme
Science and Engineering Discovery Environment (XSEDE), which is
supported by National Science Foundation grant number OCI-1053575.

\appendix

\section{Dispersion Relation and Modified Waves} \label{sec_dispersion}
The test in \S\ref{sec_linwaves} is a convergence test to 
the solution of the linearized version of the modified MHD equations that Phurbas solves.
We derive here the dispersion relation and solutions in the cases used
in the convergence test. 
We start with the MHD mass, momentum, and induction equations, with a bulk viscosity term $\rho\nabla(\zeta \nabla\cdot\mathbf{V})$  added to the momentum equation,
\begin{align}
\partial_t \rho + \rho \nabla\cdot \mathbf{V} &= 0\\
\rho \partial_t \mathbf{V} +\nabla P - (\nabla \times \mathbf{B})\mathbf{B} +\rho\nabla(\zeta \nabla\cdot\mathbf{V})&= 0\\
\partial_t \mathbf{B} + \nabla \times (\mathbf{V} \times\mathbf{B}) &= 0,
\end{align}
and for adiabatic gas we also use an energy equation
\begin{align}
\partial_t\left(\frac{P}{\rho^\gamma}\right) = 0.
\end{align}
In these equations, $\rho$ is density, $\mathbf{V}$ is velocity, $P$
is pressure, $\mathbf{B}$ is magnetic field, and $\gamma$ is the
adiabatic index.
Linearizing, and taking the viscosity $\zeta$ as constant we obtain
\begin{align}
\partial_t \rho + \rho_0 \nabla \cdot \mathbf{V} &= 0\\
\rho_0 \partial_t \mathbf{V} + \nabla P - (\nabla \times \mathbf{B})\times\mathbf{B}_0 +\rho_0\zeta\nabla(\nabla\cdot\mathbf{V}) &= 0\\
-\partial_t \mathbf{B} + \nabla\times (\mathbf{V}\times\mathbf{B}) &= 0\\
\partial_t \left(\frac{P}{P_0} - \frac{\gamma \rho}{\rho_0}\right) &= 0
\end{align}
where subscripts $0$ indicate background values, and unsubscripted variables are fluctuations.
Substituting the form $\exp(i(\mathbf{k}\cdot\mathbf{r} - \omega t))$ as a dependence for all variables 
to search for wave solutions gives
\begin{align}
-\omega \rho + \rho_0 \mathbf{k}\cdot\mathbf{V} &= 0\\
-\omega \rho_0 \mathbf{V} +\mathbf{k}P -(\mathbf{k} \times \mathbf{B})\times \mathbf{B}_0 
+\zeta i\omega \mathbf{k}(\mathbf{k}\cdot\mathbf{V}) &= 0\\
\omega\mathbf{B} + \mathbf{k}\times (\mathbf{V}\times\mathbf{B}_0) &= 0\\
-\omega \left(\frac{P}{P_0} -\frac{\gamma \rho}{\rho}\right) &= 0
\end{align}
Solving these transformed version of the mass, induction, and equation of state for the perturbed variables gives
\begin{align}
\rho &= \rho_0 \frac{\mathbf{k}\cdot \mathbf{V}}{\omega}\\
P &= \gamma P_0 \frac{\mathbf{k}\cdot\mathbf{V}}{\omega}\\
\mathbf{B} &= \frac{(\mathbf{k}\cdot\mathbf{V})\mathbf{B}_0 - (\mathbf{k}\cdot\mathbf{B}_0)\mathbf{V}}{\omega}
\end{align}
We can then substitute these into the linearized momentum equation to yield
\begin{align}
\left[ \omega^2 - \frac{(\mathbf{k}\cdot\mathbf{B}_0)^2}{\rho_0} \right] \mathbf{V} = \nonumber\\
\left \{ \left[ \frac{\gamma P_0}{\rho_0} + \frac{\mathbf{B}_0^2}{\rho_0}\right]\mathbf{k} -\frac{(\mathbf{k}\cdot \mathbf{B}_0)}{\rho_0} \mathbf{B}_0\right \} (\mathbf{k}\cdot\mathbf{V}) -\frac{(\mathbf{k}\cdot\mathbf{B}_0)(\mathbf{V}\cdot\mathbf{B}_0)}{\rho_0}\mathbf{k}
-\zeta i\omega \mathbf{k}(\mathbf{k}\cdot\mathbf{V})
\end{align}
Substituting in the Alfv\'{e}n and sound speeds
\begin{align}
\mathbf{V}_A = \sqrt{\frac{\mathbf{B}_0^2}{\rho_0}}\hat{\mathbf{B}}_0, \qquad
V_S = \sqrt{\frac{\gamma P_0}{\rho_0}}
\end{align}
we obtain
\begin{align}
\left[ \omega^2 -(\mathbf{k}\cdot\mathbf{V}_A)^2\right] \mathbf{V} = &\nonumber\\
\left \{ (V_s^2+V_A^2)\mathbf{k}
-(\mathbf{k}\cdot\mathbf{V}_A)\mathbf{V}_A\right \}(\mathbf{k}\cdot\mathbf{V}) -(\mathbf{k}\cdot\mathbf{V}_A)(\mathbf{V}\cdot\mathbf{V}_A)\mathbf{k}
-\zeta i\omega \mathbf{k}(\mathbf{k}\cdot\mathbf{V}).
\end{align}
We can now choose a propagation direction of the wave, the direction of the wave vector $\mathrm{k}$.
The wave vector $\mathrm{k}$ is chosen to lie in the $x-z$ plane, at an angle $\theta$ clockwise from the $z$-axis.
This gives
\begin{align}
[\omega^2 -k^2 V_A^2\cos^2\theta]\mathbf{V} = &\nonumber\\
 \left \{ (V_s^2 + V_A^2)\mathbf{k}  - \hat{\mathbf{z}} |k|V_A^2\cos \theta\right \}(\mathbf{k}\cdot \mathbf{V}) - k^2 \mathbf{V}_A\cos \theta (\mathbf{V}\cdot \mathbf{V}_A)
-\zeta i\omega \mathbf{k}(\mathbf{k}\cdot\mathbf{V}).
\end{align}
Upon transforming this into a matrix eigenvalue problem, and solving for the eigenvectors $\mathbf{V}$
one obtains expressions for the eigenvalues, which can be solved for $\omega$.
Unfortunately, we have only found analytic solutions for two special cases of the propagation direction $\theta$.
These cases are used to test the convergence of our scheme.
For propagation perpendicular to the background magnetic field  $\mathbf{B}_0$, 
$\theta = \pi/2$ and the first eigenvalue corresponds to a magnetoacoustic wave, with speed given by
 $\mathfrak{Re}(\omega)/k = \sqrt{|k^2\zeta^2/4 -(V_A^2+V_s^2)|}$, and
 attenuation given by
 $\mathfrak{Im}(\omega) = i k^2\zeta/2$.
For propagation in the direction of the background magnetic field
$\mathbf{B}_0$, $\theta = 0$ and the first eigenvalue corresponds to a
sound wave, with speed given by 
 $\mathfrak{Re}(\omega)/k = \sqrt{|k^2\zeta^2/4 -(V_s^2)|}$, 
and attenuation given by $\mathfrak{Im}(\omega) = i k^2\zeta/2$.
In general, the shear Alfv\'{e}n wave eigenmodes are unaffected by the addition of the bulk 
viscosity, as they do not involve compressive motions.

\end{document}